\documentclass[aps, prl, amsmath,amssymb,twocolumn, groupedaddress, superscriptaddress, reprint]{revtex4-2}
\usepackage{graphicx}
\usepackage{dcolumn}
\usepackage[utf8]{inputenc}
\usepackage[colorlinks,linkcolor=blue,citecolor=blue,urlcolor=blue]{hyperref}
\usepackage{stix}

\usepackage[dvipsnames]{xcolor}
\usepackage{xspace}

\usepackage{pdfpages}
\usepackage{etoolbox} 

\makeatletter
\patchcmd{\@outputpage@head}{\@ifx{\LS@rot\@undefined}{}{\LS@rot}}{}{}{}
\makeatother

\begin{document}
\title{Controlling \emph{4f} antiferromagnetic dynamics via itinerant electronic susceptibility} 

\author{Sang-Eun Lee}
\email{sang-eun@lanl.gov}
\affiliation{Department of Physical Chemistry, Fritz-Haber-Institut der Max-Planck-Gesellschaft, Faradayweg 4-6, 14195 Berlin, Germany}

\author{Yoav William Windsor}
\affiliation{Department of Physical Chemistry, Fritz-Haber-Institut der Max-Planck-Gesellschaft, Faradayweg 4-6, 14195 Berlin, Germany}

\author{Daniela Zahn}
\affiliation{Department of Physical Chemistry, Fritz-Haber-Institut der Max-Planck-Gesellschaft, Faradayweg 4-6, 14195 Berlin, Germany}

\author{Alexej Kraiker}
\affiliation{Physikalisches Institut, Goethe-Universität Frankfurt, Max-von-Laue-Str. 1, 60438 Frankfurt am Main, Germany}

\author{Kurt Kummer}
\affiliation{European Synchrotron Radiation Facility, BP 220, F-38043 Grenoble Cedex, France.}

\author{Kristin~Kliemt}
\affiliation{Physikalisches Institut, Goethe-Universität Frankfurt, Max-von-Laue-Str. 1, 60438 Frankfurt am Main, Germany}

\author{Cornelius Krellner}
\affiliation{Physikalisches Institut, Goethe-Universität Frankfurt, Max-von-Laue-Str. 1, 60438 Frankfurt am Main, Germany}

\author{Christian Sch{\"u}{\ss}ler-Langeheine}
\affiliation{Helmholtz-Zentrum Berlin für Materialien und Energie GmbH, Albert-Einstein-Str. 15, 12489 Berlin, Germany}

\author{Niko~Pontius}
\affiliation{Helmholtz-Zentrum Berlin für Materialien und Energie GmbH, Albert-Einstein-Str. 15, 12489 Berlin, Germany}

\author{Urs Staub}
\affiliation{Swiss Light Source, Paul Scherrer Institut, Forschungsstr. 111, 5232 Villigen PSI, Switzerland}

\author{Denis V. Vyalikh}
\affiliation{Donostia International Physics Center (DIPC), Paseo Manuel de Lardizabal, 4, 20018 Donostia/San Sebastián, Basque Country, Spain}
\affiliation{IKERBASQUE, Basque Foundation for Science, Plaza Euskadi 5, 48009 Bilbao, Spain}

\author{Arthur Ernst}
\email{arthur.ernst@jku.at}
\affiliation{Institute for Theoretical Physics, Johannes Kepler University, Altenberger Str. 69, 4040 Linz, Austria}
\affiliation{Max-Planck-Institut für Mikrostrukturphysik, Weinberg 2, 06120 Halle (Saale), Germany}

\author{Laurenz Rettig}
\email{rettig@fhi-berlin.mpg.de}
\affiliation{Department of Physical Chemistry, Fritz-Haber-Institut der Max-Planck-Gesellschaft, Faradayweg 4-6, 14195 Berlin, Germany}

\date{\today}


\begin{abstract}
Optical manipulation of magnetism holds promise for future ultrafast spintronics, especially with lanthanides and their huge, localized 4f magnetic moments. These moments interact indirectly via the conduction electrons (RKKY exchange), influenced by interatomic orbital overlap, and the conduction electron’s susceptibility. Here, we study this influence in a series of 4f antiferromagnets, Gd\emph{T}$_2$Si$_2$ (\emph{T}=Co, Rh, Ir), using ultrafast resonant X-ray diffraction. We observe a twofold increase in ultrafast angular momentum transfer between the materials, originating from modifications in the conduction electron susceptibility, as confirmed by first-principles calculations.
\end{abstract}

\maketitle

\begin{figure}
    \includegraphics[width=\linewidth]{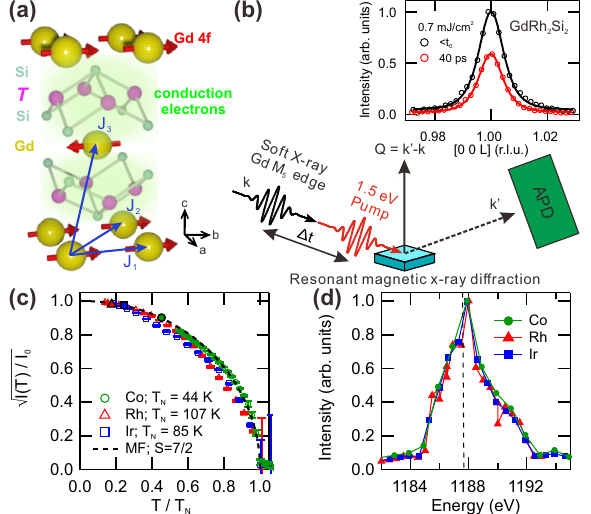}
    \caption{(a) Crystal structure of Gd\emph{T}$_2$Si$_2$ (\emph{T} = Co, Rh, and Ir). $J_1$, $J_2$ are exchange coupling between the nearest and the next nearest in-plane \emph{4f} moments, and $J_3$ is the exchange coupling between the nearest out-of-plane \emph{4f} moments. (b) Sketch of the experimental setup. (top) Pump-induced suppression of the (001) magnetic diffraction peak of GdRh$_2$Si$_2$. (c) Equilibrium temperature-dependent behavior of the magnetic diffraction amplitude of Gd\emph{T}$_2$Si$_2$ (symbols). Dashed line is a mean-field curve for $S=7/2$ corresponding to Gd~\cite{aharoni_introduction_2000}. Bold solid symbols indicate the estimated staggered magnetization of each sample at 20 K, the base temperature of the dynamic measurements. (d) Photon-energy-dependent resonant amplification of the magnetic diffraction intensity of Gd\emph{T}$_2$Si$_2$ (symbols). Vertical dashed line indicates the photon energy we chose for the rest of this study (1187.5 eV).}
    \label{F1}
\end{figure}

Lanthanides are becoming increasingly important in technology due to their exceptionally large magnetic moments. Since the majority of the magnetic moments of lanthanides resides in spatially localized \emph{4f} electron shells~\cite{aharoni_introduction_2000, stohr_magnetism_2006}, magnetic moments from different lanthanide atoms can only interact indirectly by spin-polarizing itinerant conduction electrons that surround the \emph{4f} moments. This indirect interaction is called the Rudermann-Kittel-Kasuya-Yosida (RKKY) exchange interaction \cite{jensen_mackintosh_rare_1991}. As its mechanism implies, not only the localized \emph{4f} moments but also the itinerant conduction electrons play an important role in determining the strength of the RKKY interaction $J_{RKKY}$ in lanthanide-based magnetic materials. $J_{RKKY}$ is proportional to the squared overlap integral between the \emph{4f} and conduction electrons, $\lvert I \rvert^2$, and the susceptibility $\chi$ of the conduction electrons' spin polarization around the Fermi level to an effective magnetic field formed by \emph{4f} magnetic moments $\left(J_{RKKY} \propto \lvert I \rvert^2 \chi\right)$~\cite{jensen_mackintosh_rare_1991}. 

The ultrafast dynamics of magnetic devices after femtosecond laser excitation is governed by the transfer speed of angular momentum between its microscopic subsystems. While the investigation of such ultrafast spin dynamics has been employed to study these interactions for several decades~\cite{vaterlaus_spin-lattice_1991, beaurepaire_ultrafast_1996}, and also numerous studies of lanthanide magnetism have been reported \cite{wietstruk_hot-electron-driven_2011,andres_separating_2015,frietsch_disparate_2015,rettig_itinerant_2016,thielemann-kuhn_ultrafast_2017,windsor_deterministic_2020,frietsch_role_2020, andres_role_2021,windsor_exchange_2022}, the influence of the individual contributions of $J_{RKKY}$ to ultrafast spin dynamics remained mostly elusive. In particular, whereas the role of the localized \emph{4f} magnetic moments has been extensively studied mostly in the heavy trivalent lanthanides (Gd - Tm)~\cite{wietstruk_hot-electron-driven_2011, frietsch_role_2020, andres_role_2021}, the role of the itinerant conduction electrons in lanthanide magnetism has not been investigated systematically so far since many elemental lanthanides share very similar conduction electron structures making it difficult to isolate their specific role in ultrafast magnetization dynamics. However, due to their central role in the RKKY interaction, a systematic investigation of the influence of the conduction electrons properties on the magnetization dynamics in lanthanide-based compounds is of strong interest. 

For this purpose, we investigated the ultrafast magnetization dynamics in a series of \emph{4f} antiferromagnetic (AF) compounds \emph{Ln}\emph{T}$_2$Si$_2$ (\emph{Ln}: lanthanides; \emph{T}: transition metals, Fig.~\ref{F1}a), which share almost identical magnetic and lattice structures~\cite{kliemt_single_2015, kliemt_crystal_2020}. Using this similarity, we recently investigated the role of the \emph{4f} moments on the ultrafast spin dynamics of \emph{Ln}Rh$_2$Si$_2$ by varying the lanthanide \emph{Ln} \cite{windsor_exchange_2022}, which demonstrated that the direct spin transfer between antiferromagnetically coupled \emph{4f} moments is an essential demagnetization channel during ultrafast spin dynamics scaling with the strength of the RKKY interaction. In this Letter, in a similar approach we systematically vary the nonmagnetic transition metal \emph{T} occupation within Gd\emph{T}$_2$Si$_2$ from \emph{3d} to \emph{5d} (Co, Rh, Ir), and single out the influence on the RKKY interaction and the ultrafast spin dynamics in these \emph{4f} antiferromagnets. Surprisingly, we find a non-monotonous variation of angular momentum transfer rates with d-shell occupation, with GdRh$_2$Si$_2$ showing larger transfer rates as the other two compounds. Using ab-initio calculations, we explain this behavior by the variation of the conduction electron susceptibility $\chi$ due to a competition of the $T$-ion $d$-orbital extension and the $d$-level energy splitting.

The family of intermetallics Gd\emph{T}$_2$Si$_2$ (T = Co \emph{3d}, Rh \emph{4d} and Ir \emph{5d}) crystallizes in the tetragonal ThCr$_2$Si$_2$ structure (a = b $\sim$ 4 \AA, c $\sim$ 10 \AA) and are A-type antiferromagnets, where antiferromagnetically ordered Gd ions are separated by \emph{T}$_2$Si$_2$ blocks along the c-axis (Fig.~\ref{F1}a) \cite{czjzek_study_1989, kliemt_crystal_2020}. The sample growth condition and characterization are elaborated in Supplementary Information~I. The N\'eel temperatures $T_N$ of the three samples are 45~K (Co), 107~K (Rh), and 85~K (Ir)~\cite{tung_study_1997, windsor_deterministic_2020}. Employing resonant soft X-ray diffraction (RXD), we measured the resonantly enhanced [0~0~$L$] magnetic diffraction intensity, sensitive to long-range AF ordering of Gd \emph{4f} moments along the c axis (Fig.~\ref{F1}b). The samples were characterized at the RESOXS end station of the SIM beamline of the Swiss Light Source at the Paul Scherrer Institute, Switzerland and the RIXS end station of the ID32 beamline of the European Synchrotron Radiation Facility in Grenoble, France~\cite{flechsig_performance_2010, brookes_beamline_2018}. Time-resolved resonant soft X-ray diffraction (trRXD) experiments were performed at the FemtoSpeX beamline UE56/1-ZPM of BESSY II of the Helmholtz-Zentrum Berlin, Germany, which uses femtosecond slicing to provide ultrashort soft X-ray pulses~\cite{holldack_femtospex_2014}. We used 50 fs-long laser pulses centered at 1.55 eV, at a repetition rate of 3 kHz to excite the sample, and measured the transient diffraction intensity with 100 fs-long sliced soft X-ray pulses centered at the Gd M$_5$ absorption edge with an avalanche photodiode (APD), at a repetition rate of 6 kHz (Fig.~\ref{F1}b). All dynamical experiments were conducted at a temperature of 20~K. 

While GdIr$_2$Si$_2$ shows a commensurate magnetic diffraction peak at constant $L$ = 1 at all temperatures (see Supplementary Information~II) similar to GdRh$_2$Si$_2$~\cite{windsor_deterministic_2020}, GdCo$_2$Si$_2$ displays incommensurate magnetic diffraction peaks at $L$=$q$ and $2-q$ with $q\sim0.966$ at 20 K, which shift with temperature (see Supplementary Information~II). Since the two incommensurate peaks exhibit an almost identical temperature- and photon-energy dependence, we concentrate on the (0~0~$q$) peak in this study. 

The magnetic diffraction intensity of the three samples exhibits almost identical temperature dependencies following a mean-field-like behavior (Gd, $S=7/2$) (Fig.~\ref{F1}c). The similar photon energy dependence of resonant enhancement at the Gd M$_5$ absorption edge (h$\nu$ = 1.18 keV; \emph{3d} $\rightarrow$ \emph{4f}) demonstrates their similar orbital and magnetic configuration (Fig.~\ref{F1}d), and leads to very similar penetration depths of $\sim$4 nm at resonance, corresponding to the X-ray light probes $\sim$4 unit cells of Gd\emph{T}$_2$Si$_2$~\cite{windsor_deterministic_2020, windsor_exchange_2022, lee_systematic_2023}. These similarities in both temperature- and photon-energy dependence demonstrate their similar long-range \emph{4f} antiferromagnetism, justifying the following comparative analysis of ultrafast spin dynamics.

\begin{figure}
    \includegraphics[width=\linewidth]{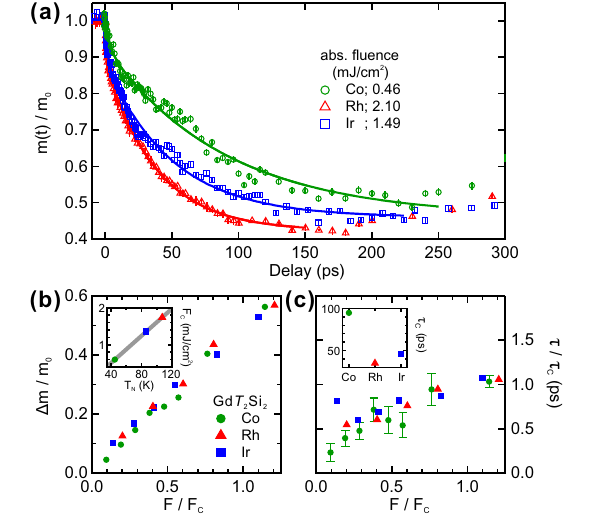}
    \caption{(a) Ultrafast dynamics of the normalized (0~0~$L$) magnetic diffraction amplitude of Gd\emph{T}$_2$Si$_2$ (\emph{T} = Co, Rh, and Ir) at a selected pump fluence acquired at a constant momentum transfer $Q$. Error bars denote uncertainties from Poisson statistics. Solid lines are exponential decaying functions for phenomenological description (see text). (b) Demagnetization amplitude of the three samples plotted along the fluence $F$ normalized by the critical fluence $F_C$ (see text) of each sample. (Inset) The relation between the critical fluence and the N\'eel temperature of each sample. (c) Normalized slow demagnetization time constants $\tau/\tau_C$ of the three samples plotted along the normalized fluence $F/F_C$. (Inset) Time constants $\tau_C$ at the critical fluence used for normalization.}
    \label{F2}
\end{figure}

Femtosecond dynamics of the (0~0~$L$) magnetic diffraction peak amplitude for selected pump fluences are shown in Fig.~\ref{F2}a. The dynamics at all measured pump fluences discussed in this study are presented in Supplementary Information~III. In the case of GdRh$_2$Si$_2$, the peak amplitudes have been separated from a transient reorientation of the magnetic in-plane easy-axis using a procedure that combines several azimuthal orientations~\cite{windsor_deterministic_2020}. To account for pump-induced transient peak shifts, the peak amplitudes for GdCo$_2$Si$_2$ have been corrected by considering the transient peak position evolution of the (0~0~$L$) diffraction peak, as detailed in Supplementary Information~IV. 

For a quantitative comparison of the demagnetization dynamics of the three materials, we modeled the demagnetization curves using a phenomenological exponential decaying functions describing different timescales:
\begin{equation}
A(t)= 1-\Theta\left(t,t_0\right)\sum_{i}^{f, s}  A_{i} \left(1-e^{-\left(t-t_0\right)/\tau_{i}}\right),  
\end{equation}
where $A_{f}$, $A_{s}$, and $\tau_{f}$, $\tau_{s}$ are the amplitude and the time constant of the fast ($\sim$1~ps) and slow ($\sim$30-100 ps) processes, respectively. $t_0$ corresponds to the temporal overlap of pump and probe pulses, and $\Theta\left(t,t_0\right)$ is the Heaviside function. Fig.~\ref{F2}b presents the demagnetization amplitude $\Delta m=A_{f}+A_{s}$ as functions of fluence normalized to the critical fluence $F_C$, defined for each material as the fluence necessary to induce 50\% demagnetization~\cite{windsor_exchange_2022} (Details of the absorbed fluence estimation are provided in Supplementary Information~V). We find values for $F_C$ of 0.60 (Co), 1.74 (Rh), and 1.36 (Ir) mJ/cm$^2$, respectively, which surprisingly do not follow the $d$-shell occupation of the $T$-ions, but instead show a linear relationship to $T_N$ (inset of Fig.~\ref{F2}b). A similar relation with a comparable slope has also been observed in\emph{Ln}Rh$_2$Si$_2$ (\emph{Ln}=Pr - Ho)~\cite{windsor_exchange_2022}. This scaling relation between the critical fluence $F_C$ and the N\'eel temperature $T_N$ implies that the relevant magnetic interactions follow a classical mean-field-like behavior~\cite{windsor_exchange_2022}, supporting the validity of our comparative analysis of the ultrafast spin dynamics of Gd\emph{T}$_2$Si$_2$.

The normalized time constants of the slow demagnetization process $\tau_{s}$, which is present in all materials, is shown in Fig.~\ref{F2}c. They exhibit a qualitatively similar square-root-like behavior with $F/F_C$, albeit the absolute time scales differ substantially (see inset of Fig.~\ref{F2}c), similar to our previous study~\cite{windsor_exchange_2022}.
Thus, while all of the studied materials exhibit qualitatively similar demagnetization behavior, there are also notable differences. Whereas GdRh$_2$Si$_2$ and GdIr$_2$Si$_2$ display a two-step decay in their demagnetization dynamics ($\tau_{f}\sim$1 ps, $\tau_{s}$>10ps), followed by a slow recovery after $\sim$100 ps, $\tau_{f}$ is almost absent in GdCo$_2$Si$_2$. Furthermore, similar to the critical fluences, we find substantial differences in the demagnetization rate, following the same sequence (GdRh$_2$Si$_2$ and GdCo$_2$Si$_2$ exhibiting the fastest and the slowest dynamics, respectively).

For accurate comparison of the demagnetization rates, the ultrafast angular momentum transfer rate $\alpha = m_{stag.} \mu_{4f} A_{s}/\tau_{s}$ is calculated for each material \cite{windsor_exchange_2022}, where $\mu_{4f}$=7$\mu_B$ is the size of the Gd \emph{4f} moments~\cite{jensen_mackintosh_rare_1991}. The staggered magnetization $m_{stag.}$ of each sample at 20 K is indicated in Fig.~\ref{F1}c with bold solid markers. The fluence-dependent behavior of the angular momentum transfer rates is shown in Supplementary Information~III. Similarly to the inverse time constants and critical fluences, GdRh$_2$Si$_2$ (4$d$) has the largest $\alpha$ followed by GdIr$_2$Si$_2$ (5$d$), and GdCo$_2$Si$_2$ (3$d$). Although they share the same Gd \emph{4f} moments, the angular momentum transfer rate of the Gd\emph{T}$_2$Si$_2$ series varies by $\sim$100\%.


\begin{figure}
    \includegraphics[width=\linewidth]{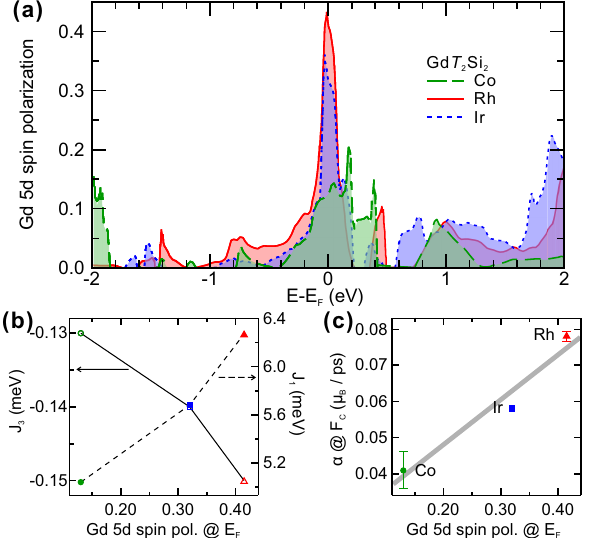}
    \caption{(a) Energy distribution of calculated spin polarization of Gd \emph{5d} states of Gd\emph{T}$_2$Si$_2$ in the vicinity of the Fermi level $E_F$. (b) Calculated indirect RKKY exchange interaction between the nearest in-plane \emph{4f} moments $J_1$ (closed markers, dashed line, right axis) and the nearest out-of-plane antiferromagnetically coupled \emph{4f} moments $J_3$ (open markers, solid line, left axis) plotted along the calculated Gd \emph{5d} spin polarization. (c) Angular momentum transfer rate at $F_C$ interpolated from fluence dependence (Fig.~III.1.d in Supplementary Information~III) plotted along the Gd \emph{5d} spin polarization of the three samples. Grey line is a guide to the eyes. The error bars are derived from error propagation of demagnetization time constant and amplitude of each compound.}
    \label{F3}
\end{figure}

In order to understand the reason for the varying ultrafast angular momentum transfer rate of the Gd\emph{T}$_2$Si$_2$ series, we calculated exchange coupling constants and electronic densities of states (eDOS) of Gd\emph{T}$_2$Si$_2$ employing density functional theory (DFT) (see Supplementary Information~VI for details). According to the DFT calculations, in Gd\emph{T}$_2$Si$_2$, the conduction electrons are composed mostly of \emph{5d} electrons. Thus, the interaction between the local magnetic moments is mediated predominantly via spin-polarized Gd \emph{5d} states. As discussed in the introduction, the strength of the RKKY interaction $J_{RKKY}$ is determined by the overlap integral between \emph{4f} and conduction electrons and the non-local susceptibility of the conduction electrons' spin polarization around the Fermi level $E_F$. Since all the studied compounds share the same local magnetic moments and Gd \emph{5d} electrons, the overlap integral factor does not vary much among the Gd\emph{T}$_2$Si$_2$ series. 

In contrast, the calculations show that the spin polarization of Gd \emph{5d} electrons around $E_F$ i.e., the difference in eDOS of majority- and minority-spin states, varies substantially between the different compounds (Fig.~\ref{F3}a). At the Fermi level $E_F$, GdRh$_2$Si$_2$ has the largest spin polarization, followed by GdIr$_2$Si$_2$ and GdCo$_2$Si$_2$. As shown in Fig.~\ref{F3}b, the spin polarization at $E_F$ also directly correlates with the strength of the calculated RKKY interaction between the nearest in-plane and out-of-plane Gd \emph{4f} moments ($J_1$, $J_3$, respectively in Fig.~\ref{F1}a). This varying spin polarization implies that the transition metal ions modify the eDOS of both majority- and minority-spin states and thus the non-local susceptibility of the conduction electrons' spin polarization in the vicinity of the Fermi level. At the same time, we also find a clear scaling relation of the experimental angular momentum transfer rate with the Gd \emph{5d} spin polarization at $E_F$ (Fig.~\ref{F3}c). As shown for the \emph{Ln}Rh$_2$Si$_2$ series~\cite{windsor_exchange_2022}, the ultrafast angular momentum transfer rate in this series of compounds scales with the strength of the RKKY interaction. Therefore, the observed scaling relation with the Gd $5d$ spin polarization (Fig.~\ref{F3}c) reflects the influence of the conduction electrons' susceptibility on the ultrafast spin dynamics of the Gd\emph{T}$_2$Si$_2$ series due to varying nonmagnetic \emph{T} ions.

\begin{figure}
    \includegraphics[width=\linewidth]{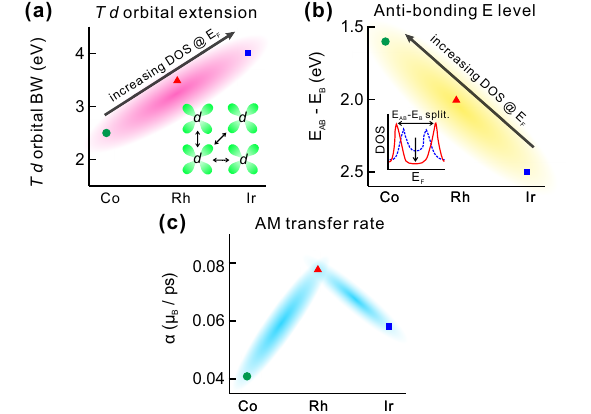}
    \caption{Cartoon summarizing the factors contributing to the scaling of the ultrafast angular momentum transfer rate of Gd\emph{T}$_2$Si$_2$ with respect to the transition metal (\emph{T}) ions, \emph{3d}~Co, \emph{4d}~Rh and \emph{5d}~Ir. (a) The extension of \emph{T} \emph{d}-orbitals is reflected by their calculated bandwidth (Fig.~VII.1 in Supplementary Information~VII). (b) The distance between the bonding~/~antibonding orbitals with respect to the Fermi level is reflected by the energy splitting of the calculated unoccupied antibonding \emph{T d} and occupied \emph{T d} states (Fig.~Fig.~VI.1). (c) The combined effect of the trends in (a) and (b) leads to the observed behavior of spin susceptibility and angular momentum transfer rates.}
    \label{F4}
\end{figure}

The behaviour of \emph{5d} electrons in Gd\emph{T}$_2$Si$_2$ explored in our calculations can be explained by two important factors (Fig.~\ref{F4}). The first important factor is the extension of the transition metal wave functions. These orbitals show an increasing degree of delocalization when going from the \emph{3d} to the \emph{5d} shell, with GdCo$_2$Si$_2$ showing the strongest localization (see Supplementary Information~VII). Therefore, the hybridization between Si and Co states is much weaker than the hybridization between Si and Ir~/~Rh states in GdIr$_2$Si$_2$ and GdRh$_2$Si$_2$. Consequently, with increasing delocalization of \emph{T} $d$ orbitals along the series the vacant Si valence electrons hybridize less with the Gd~\emph{5d} states, increasing the \emph{5d} eDOS and hence the spin polarisation at the Fermi level (Fig.~\ref{F4}a). The second factor influencing the \emph{5d} DOS is the bonding~/~antibonding splitting of \emph{T} $d$ states, and their distance to $E_F$. Here, the distance of the antibonding states from $E_F$ increases from Co to Ir (see Supplementary Information~VII), leading to a reduction of the eDOS near $E_F$ along the series (Fig.~\ref{F4}b), which in consequence decreases the spin polarization near $E_F$ as well. Combined with the first factor, this explains the observed behavior the largest eDOS and angular momentum transfer in GdRh$_2$Si$_2$ (Fig.~\ref{F4}c). 

In addition, the particular crystalline structure of Gd\emph{T}$_2$Si$_2$ supports this trend. As shown in Ref.~\cite{Hughes2007}, magnetic properties of lanthanide compounds are highly sensitive to changes in unit cell volumes: a reduction of the unit cell volume leads to a reduction of \emph{5d} eDOS at the Fermi level, thereby modifying the magnetic interaction in the system. In our case, the variation of the unit cell volume (GdCo$_2$Si$_2$: 150.0~\AA$^3$~\cite{czjzek_study_1989} < GdIr$_2$Si$_2$: 156.0~\AA$^3$~\cite{kliemt_crystal_2020} < GdRh$_2$Si$_2$: 162.9~\AA$^3$~\cite{kliemt_single_2015}) reflects the changes in eDOS and spin polarization at $E_F$.

In summary, we investigated the role of the itinerant conduction electrons in ultrafast spin dynamics of \emph{4f} antiferromagnets. By substituting the \emph{T} ions in Gd\emph{T}$_2$Si$_2$ (\emph{T}= Co, Rh, Ir), we selectively modified their conduction electron susceptibility and measured femtosecond dynamics of magnetic diffraction intensity at various pump fluences employing time-resolved resonant magnetic soft x-ray diffraction. While we found qualitatively similar demagnetization behavior upon optical excitation at 1.55 eV, the observed critical fluences and ultrafast angular momentum transfer rates $\alpha$ vary drastically, and non-monotonously with $T$ orbital shell ($\alpha$: GdRh$_2$Si$_2$ $>$ GdIr$_2$Si$_2$ $>$ GdCo$_2$Si$_2$). First-principles calculations of electronic density of states and exchange coupling constants of Gd\emph{T}$_2$Si$_2$ employing density functional theory show that the spin polarization of Gd \emph{5d} electrons scales with the in-plane and out-of-plane nearest neighbor exchange coupling constants, and with the experimental angular momentum transfer rate. This implies that varying the \emph{T} ions modifies the non-local susceptibility of conduction electrons' spin polarization and hence, the strength of the RKKY interaction. We explain this effect by a combination of \emph{d} orbital wavefunction localization and bonding~/~antibonding splitting of \emph{T d} states, modifying the electronic density of states around the Fermi level and their non-local susceptibility. Our findings provide important insights for designing lanthanide-based magnetic devices, showing how a modification of the itinerant conduction electrons, which could e.g. be implemented by chemical or electrostatic doping, impacts ultrafast angular momentum transfer processes.

\section*{Acknowledgements} 
We thank the Helmholtz-Zentrum Berlin für Materialien und Energie for the allocation of synchrotron radiation beamtime. The experimental support of the staff at beamlines UE56/1 (HZB), X11MA (SLS), and PM3 (HZB) is gratefully acknowledged. This work received funding from the Deutsche Forschungsgemeinschaft (DFG, German Research Foundation) within the Emmy Noether program (Grant No. RE 3977/1), within the Transregio TRR 227 - 328545488 Ultrafast Spin Dynamics (Projects A03 and A09), within TRR 288 - 422213477 (Project A03), SFB1143 (No. 247310070) and within Grant No. 282 KR3831/5-1. Funding was also received from the European Research Council (ERC) under the European Union’s Horizon 2020 research and innovation program (Grant Agreement Number ERC-2015-CoG-682843). We also thank support from the Spanish Ministry of Science and Innovation, project PID2020-116093RB-C44, funded by MCIN/ AEI/10.13039/501100011033. We acknowledge funding by Fonds zur Förderung der wissenschaftlichen Forschung (FWF) grant I 5384. Calculations were carried out at the Rechenzentrum Garching of the Max-Planck Society. We acknowledge the European Synchrotron Radiation Facility (ESRF) for provision of synchrotron radiation facilities under proposal number IH-HC-3815.



\section*{Supplementary material}

## I  Crystal growth and characterization

The $GdRh_2Si_2$ and $GdIr_2Si_2$ single crystals were obtained according to the procedure described earlier [1, 2]. For the growth of $GdCo_2Si_2$ single crystals, high-purity starting materials were weighed in a graphite crucible and sealed in a niobium crucible under argon atmosphere. The stoichiometric mixture of the elements was used with the ratio 1 : 2 : 2 : 24 (Gd : Co : Si : In) with indium as flux. The growth was performed as described in Ref. [2] with a maximum temperature of the furnace of 1550 °C.

Powder x-ray diffraction yielded lattice parameters of $a = b = 3.911$Å and $c = 9.803$Å which are in good agreement with literature [3]. Energy dispersive x-ray spectroscopy (EDX) revealed the stoichiometry of Gd : Co : Si = $20 \pm 1$ : $39 \pm 2$ : $41 \pm 2$ which is in good agreement with the 122 target stoichiometry within the experimental error. The orientation of the single crystals was determined using the Laue method.

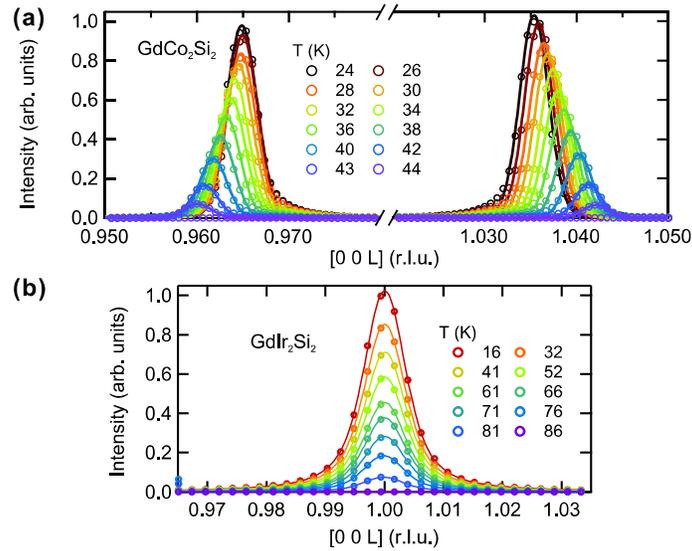

FIG. II.1. Temperature dependent behaviour of the magnetic diffraction intensity of (a) $GdCo_2Si_2$ and (b) $GdIr_2Si_2$.

---

[*] sang-eun@lanl.gov

[†] arthur.ernst@jku.at

[‡] rettig@fhi-berlin.mpg.de

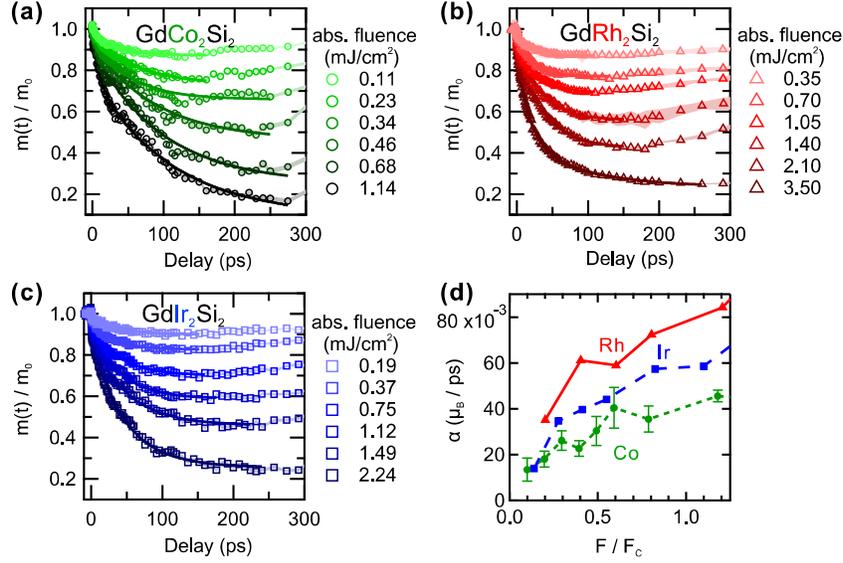

FIG. III.1. (a-c) Ultrafast antiferromagnetic order dynamics of Gd$T_2$Si$_2$ ($T$= Co, Rh, Ir) at various pump fluences. Solid lines are phenomenological description of the decaying part of the curves using exponentially decaying functions (see main text). (d) Ultrafast angular momentum transfer of Gd$T_2$Si$_2$ plotted along the normalized fluence (fluence divided by the critical fluence of each material; see main text).

## II  Equilibrium temperature dependence of the magnetic diffraction intensity of GdIr$_2$Si$_2$ and GdCo$_2$Si$_2$

Here we discuss the equilibrium temperature-dependent behavior of the magnetic diffraction peak of Gd$T_2$Si$_2$. The case of GdRh$_2$Si$_2$ is presented in Ref. [4]. The cases of GdCo$_2$Si$_2$ and GdIr$_2$Si$_2$ are presented in Fig. II.1a/b, respectively. While GdIr$_2$Si$_2$ exhibits a commensurate diffraction peak at constant $Q$= 1 at all temperatures as GdRh$_2$Si$_2$ [4], GdCo$_2$Si$_2$ exhibits two incommensurate diffraction peaks at [0 0 $q$] and [0 0 2 − $q$] where $q \sim 0.966$ at 24 K. The peak position shifts towards smaller $q$ with increasing temperature.

## III  Ultrafast antiferromagnetic order dynamics of Gd$T_2$Si$_2$

Ultrafast long-range *4f* sublattice magnetization dynamics of Gd$T_2$Si$_2$ ($T$= Co, Rh, Ir) at various pump fluences are presented in Fig. III.1a-c. Curves presented in Fig. 2 of the main text are selected from this set of demagnetization curves in Fig. III.1. As discussed in the main text, note that the

3

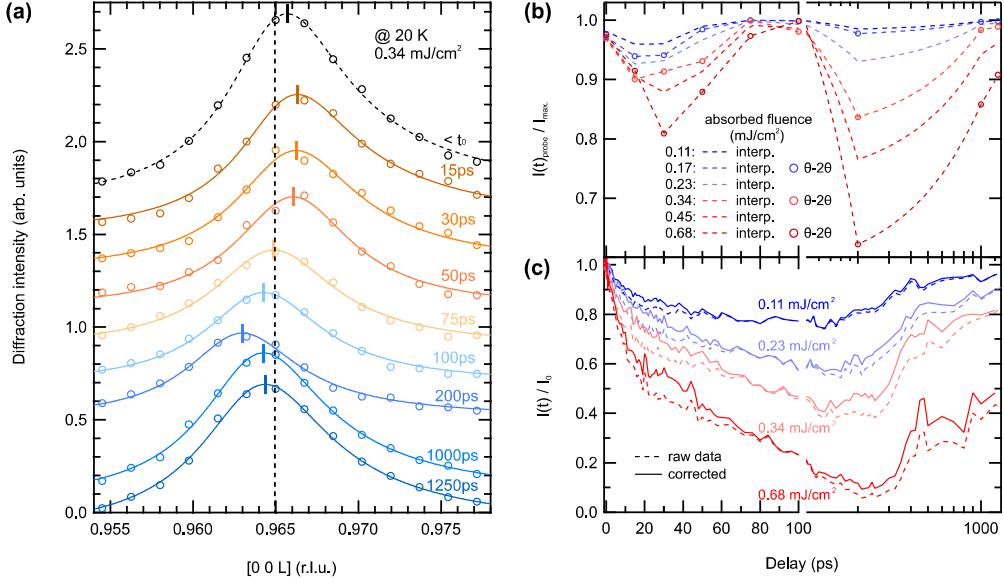

FIG. IV.1. (a) Transient diffraction intensity evolution of the (0 0 $q$) reflection of $GdCo_2Si_2$ upon ultrafast optical excitation at selected delays. The vertical dashed line indicates the diffraction geometry for acquiring delay scans (see text). (b) The ratio between the intensity for probing the delay scan $I_{probe}$ and the actual maximum intensity of the (0 0 $q$) diffraction intensity $I_{max}$ at various pump fluences. Dashed lines indicate the linearly interpolated ratio based on the experimental data points. (c) The raw diffraction intensity dynamics (dashed lines) and the corrected diffraction intensity dynamics (solid lines) at various pump fluences.

demagnetization dynamics of $GdCo_2Si_2$ are corrected for a possible transient magnetic diffraction peak shift, which will be detailed in Section IV.

The fluence-dependent behavior of the ultrafast angular momentum transfer rate of $GdT_2Si_2$ upon optical excitation is plotted in Fig. III.1d. While the angular momentum transfer rate increases with the normalized fluence, the systematic difference between the investigated compounds with $GdRh_2Si_2$ having the largest rate and $GdCo_2Si_2$ the smallest transfer rate remains consistent.

## IV Correction of the demagnetization amplitude considering the transient peak shift in $GdCo_2Si_2$

Since the delay scans were acquired at constant momentum transfer $Q$, they may not reflect the correct antiferromagnetic order dynamics due to a transient peak shift in $GdCo_2Si_2$. In order to

correct possible intensity modulation from the peak shift, we measured the transient evolution of the diffraction peak position of $GdCo_2Si_2$ at selected fluences and at selected delays (Fig. IV.1a). The delay scans were acquired at the "shoulder" (the vertical dashed line) of the equilibrium peak to minimize such intensity variation. The diffraction peak intensity of $GdCo_2Si_2$ is modeled with a phenomenological Doniach-Sunjic function for precise description of the asymmetric peak shape:

$$DS(L; \alpha, \Gamma, Q, A) = A \left( \frac{\cos \frac{\pi \alpha}{2} + (1-\alpha) \tan \frac{L-Q^{-1}}{\Gamma}}{\left(\Gamma^2 + (L-Q)^2\right)^{(1-\alpha)/2}} \right), \quad (1)$$

where $L$ is the position in reciprocal space $[0\ 0\ L]$. $A$ is the amplitude of the peak, $Q$ is the effective peak position, $\Gamma$ is the effective peak width, and $\alpha$ determines the degree of asymmetry of the peak. For $\alpha = 0$, the equation becomes a symmetric Lorentzian profile, and asymmetry increases as $\alpha$ increases towards 1. The modeled intensity at the delay scan acquisition point $I_{probe}$ (vertical dashed line) is clearly different from the modeled maximum intensity of the peak $I_{max.}$ for most pump-probe delays. The ratio between the two ($I_{probe}/I_{max.}$) are plotted in Fig. IV.1b. As we acquired the diffraction peak at selected delays, delay points in between the $\theta - 2\theta$ scans was linearly interpolated. Similarly, the fluence dependence in between measured fluences was linearly interpolated using the delay-interpolated ratio that was constructed in the previous step. The interpolated correction factors shown in Fig. IV.1b were applied to the raw delay scan intensity.

Fig. IV.1c presents the corrected delay scans (solid lines) along with the raw delay scans (dashed lines) at selected fluences. As we see, the corrected intensity compensates the intensity loss from the transient peak at early and later phases of the dynamics.

## V  Estimation of absorbed fluence

All the fluences reported are total absorbed fluences, which were calculated using the measured incident fluences corrected for reflection and refraction effects. Calculation of absorbed fluence of $GdRh_2Si_2$ is reported in Ref. [4]. We conducted reflectivity measurements of the other two samples using 1.55 eV light. We estimate the complex index of refraction at this photon energy as $n = n_0 + ik$; $GdIr_2Si_2$: $(2.97 \pm 0.18) + (2.52 \pm 0.25)i$, $GdCo_2Si_2$: $(4.64 \pm 0.46) + (3.77 \pm 0.68)i$. Using the indices of refraction, the total absorbed fluences were estimated. We estimate the penetration depth of 1.55 eV light to be 15.3 nm ($GdCo_2Si_2$), 26.3 nm ($GdIr_2Si_2$) at the Bragg angle.

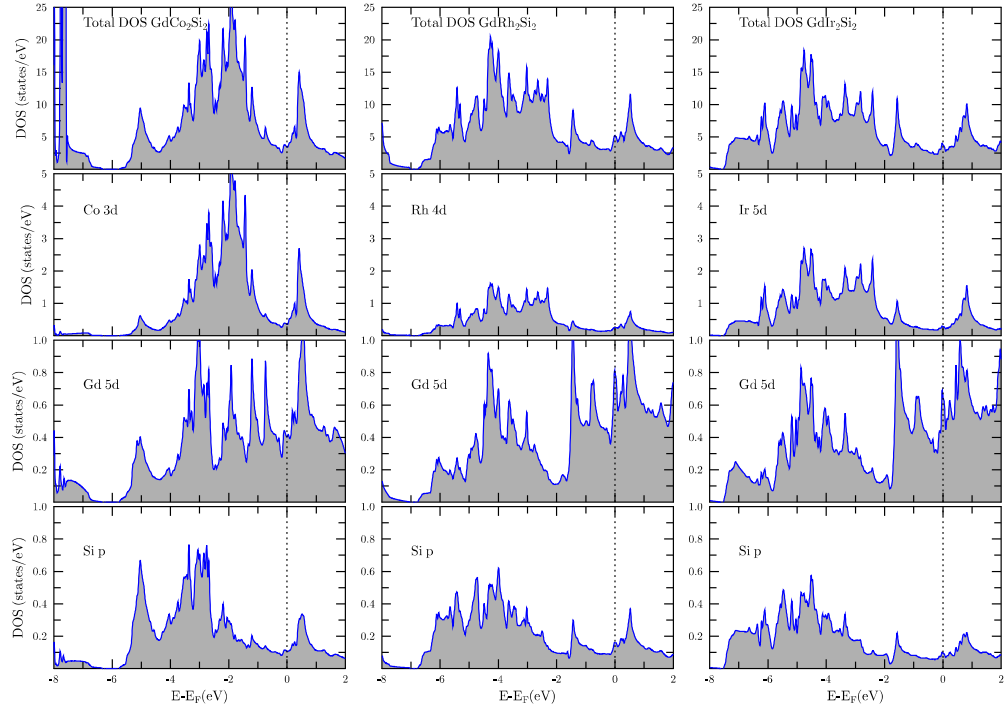

FIG. VII.1. Total and atomic resolved densities of states (DOS) of Gd$T_2$Si$_2$: $T=$ Co (left panel), Rh (middle), Ir (right panel). In the atomic resolved DOS only most important orbital contributions are shown.

## VI    First-principles calculations

First-principles calculations were carried out using a self-consistent Greens function method [5, 6] within the density functional theory in a generalized gradient approximation [7]. Strongly localized Gd *4f* electronic state were treated within a GGA+*U* method applying $U = 6.0$ eV[8]. Exchange constants were obtained utilizing the magnetic force theorem as it is implemented within the multiple scattering theory [9]. Critical temperatures were estimated within a random phase approximation [10].

## VII    Electronic structure of Gd$T_2$Si$_2$ (*T*= Co *3d*, Rh *4d*, Ir *5d*)

To elucidate the nature of the exchange coupling and observed magnetic properties in Gd$T_2$Si$_2$ (T= Co *3d*, Rh *4d*, Ir *5d*), the densities of states were calculated and analyzed (see Fig. VII.1). Gd *4f* states are localized and located mostly 8 eV below the Fermi level. These states form local-

ized magnetic moments which interact with each other via itinerant conduction electrons (RKKY interaction). Since the spin polarization of the free electrons is crucial for the RKKY interaction, mainly Gd 5*d* and Si 3*p* states participate in the magnetic interaction. Si has an induced magnetic moment: 0.08 $\mu_B$, 0.11 $\mu_B$, 0.09 $\mu_B$ for the cases with Co, Rh and Ir, respectively. Gd *5d* electrons carry moments almost two times larger than Si *3p*: 0.18 $\mu_B$, 0.21 $\mu_B$, 0.19 $\mu_B$, respectively. It is remarkable that the conduction electrons are more spin-polarized in GdRh$_2$Si$_2$ than in the other two compounds and this is in line with the strength of the calculated *J*'s presented in the main text and observed Néel temperatures.

*d* electrons of the transition metals are not spin-polarized and therefore do not participated directly in the magnetic coupling. However, they are crucial for the covalent binding in the compounds and are responsible for formation of structural and electronic properties. Co *3d* states are strongly localized (mainly between 1 and 4 eV below the Fermi level). The localized nature of Co *3d* electrons in GdCo$_2$Si$_2$ is responsible for the significantly smaller unit cell volume than in GdRh$_2$Si$_2$ and GdIr$_2$Si$_2$: both *a* and *c* lattice parameters are about 5% smaller that in the other two compounds. Despite the smaller volume, Co *3d* states remain to be localized and hybridize less with Gd *5d* and Si *3p* states compared to Rh *4d* and Ir *5d* states. The later two are more delocalized: The Rh *4d* and Ir *5d* bandwidth extends the bottom of the valence zone at 6.7 eV and 7.6 eV below the Fermi level, respectively. From Fig. VII.1 one recognizes that Si *3p* and Gd *5d* electrons hybridize with the transition metal *d* orbitals only in the lower part of the valence zone: States below -2 eV belong mainly to $d^{xz}$, $d^{3z^2-r}$ and $d^{yz}$ symmetries, while close to the Fermi level the in-plane $d^{xy}$, $d^{x^2-y^2}$ symmetries dominate. In the case of GdCo$_2$Si$_2$, Gd *5d* states with $d^{xy}$ and $d^{x^2-y^2}$ symmetries are more dispersive since the in-plane lattice constant *a* is small and this leads to a stronger hybridization within the Gd layer. In the other two compounds, out-of-plane and in-plane Gd *5d* orbitals are clearly separated. This separation and a stronger localization of the orbitals results in a larger DOS at the Fermi level in GdRh$_2$Si$_2$ and GdIr$_2$Si$_2$, which increases the strength of the RKKY interaction. However, despite the more extended Ir *5d* orbitals, the DOS at the Fermi level is larger in GdRh$_2$Si$_2$ (and thereby the RKKY interaction is stronger). The main reason for this is the position of the transition metal anti-bonding *d* states above the Fermi level: in GdRh$_2$Si$_2$ they are located at 0.5 eV above the Fermi level, which is about 0.4 eV lower in energy than in GdIr$_2$Si$_2$. This leads to a stronger accumulation of the DOS at the Fermi level. In the case of GdCo$_2$Si$_2$ this argument does not hold, since the strong in-plane hybridization of Gd *5d* electrons makes the states

7

more broad and reduces the DOS at the Fermi level as discussed above.



\end{document}